\documentstyle[titlepage,fleqn,12pt]{article}
\begin{document}
\begin{titlepage}
\title{Biharmonic pattern selection}
\author{Wei Wang$^{\dag\ddag}$ and E. Canessa$^{\dag}$\\
\\
$^{\dag}$Condensed Matter Group\\
ICTP-International Centre for Theoretical Physics\\
P.O. Box 586, 34100 Trieste, Italy\\
\\
$^{\ddag}$Physics Department; The Centre for Non-linear Dynamical Systems\\
Nanjing University, Nanjing, People's Republic of China
}

\date{}
{\baselineskip=23pt

\begin{abstract}

A new model to describe fractal growth is discussed which includes effects
due to long-range coupling between displacements $u$.  The model is
based on the biharmonic equation $\nabla^{4}u =0$ in two-dimensional isotropic
defect-free media as follows from the Kuramoto-Sivashinsky equation for
pattern formation -or, alternatively, from the theory of elasticity.  As a
difference with Laplacian and Poisson growth models, in the new
model the Laplacian of $u$ is neither zero or proportional to $u$.
Its discretization allows to reproduce a transition from dense to multibranched
growth at a point in which the growth velocity exhibits a minimum similarly to
what occurs within Poisson growth in planar geometry.
Furthermore, in circular geometry the transition point is estimated for the
simplest case from the relation $r_{\ell}\approx L/e^{1/2}$ such that the
trajectories become stable at the growing surfaces in a continuous limit.
Hence, within the biharmonic growth model, this transition depends only on
the system size $L$ and occurs approximately at a distance $60 \%$ far from a
central seed particle.  The influence of biharmonic patterns on the growth
probability for each lattice site is also analysed.

\vskip 2cm

PACS numbers: 61.50.Cj, 68.70.+w, 05.40.+j, 03.40.D
\end{abstract}
    }
\maketitle
\end{titlepage}
\baselineskip=23pt
\parskip=0pt

\section{Introduction}

The study of pattern formation in physically important fields by using
simple lattice models under certain boundary conditions is of great
interest because they describe many phenomena as occur in nature.
At present, numerical simulation studies include, {\em e.g.},
the dielectric breakdown model (DBM), {\em i.e.}, solving the Laplacian
equation in the medium surrounding the growing aggregates
\cite{Nie84,Arian89,Can91}, and the Poisson growth model to the pattern
formation in screened electrostatic fields \cite{Lou92}.
Besides, the diffusion-limited aggregation (DLA) model also plays a
crucial role in illustrating fractal growth
\cite{LaRo91,Bow91,Arne92}.  So far, many efforts have been invested in
such stochastic models that share the common feature of being essentially
second-order differential equations.
In these models iterative procedures are carried out around {\em four} mesh
points of a lattice to then phenomenologically relate the global influence
of a growing pattern to the growth probability for each lattice site under
different power-law forms (see, {\em e.g.}, \cite{Meak91}).

Nevertheless, there is also another important class of physical problems
leading to partial differential equations that can also be linear in the
order parameter but, as a peculiarity, they are of higher order
and directly concern the issue of pattern formation.  An example of this
is the (time-averaged) Kuramoto-Sivashinsky equation
\cite{Kura78,Siva79}:
\begin{equation}\label{eq:1q}
\frac{\partial u}{\partial t}= \nu \nabla^{2} u +\lambda \nabla^{2}\nabla^{2}
         u+\mu (\nabla u)^{2}  \;\;\; ,
\end{equation}
which models pattern formation in different physical contexts, such as
chemical reaction-diffusion systems \cite{Hym86,Shra86} and a cellular
gas flame in the presence of external stabilizing factors \cite{Malo84}.
In the above $\nu$, $\lambda$ and $\mu$ are normalized coefficients and $u$
a displacement at time t.

If we consider the simplest case, {\em i.e.} assume static solutions and
keep linear terms only ({\em i.e.}, $\mu \equiv 0$), then Eq.(\ref{eq:1q})
reduces to the sum of two important terms: the Laplacian equation plus the
{\em Biharmonic} equation
\begin{equation}\label{eq:w1}
\nabla^{4}u =\nabla^{2}(\nabla^{2}u )=0   \;\;\; .
\end{equation}
Since the former term ({\em i.e.} $\lambda =0$), when discretized, is known
to display fractal structures on a squared lattice \cite{Nie84,Arian89},
then the latter term ({\em i.e.}, Eq.(\ref{eq:w1})) might also in fact
select a new class of fractal patterns.

In addition to the above physical relevance, the biharmonic part of
Eq.(\ref{eq:1q}) can describe the deflection of a thin plate subjected
to uniform loading over its surface with fixed edges \cite{Can91,Land59},
the steady slow two-dimensional (2D) motion of a viscous fluid \cite{Lamb32}
or the vibration modes in the acoustic of drums \cite{Ross92}.  Besides
this a higher-order differential equation containing the biharmonic
term also appears in the study of kinetic growth with surface
relaxation (see, {\em e.g.}, \cite{Wo90,Lai91,Lam91,Yan92}).

As an important difference with respect to second-order,
to solve numerically higher-order differential equations, such as
the one at hand, requires values for
the order parameters at either their first or second normal derivatives at
each boundary point and beyond \cite{Gde86,Bue91}.  Hence, an analysis of
effects due to long-range (many-body) coupling of
lattice sites, including those mesh points at the lattice boundaries, on
the formation of connected patterns is by no means trivial and as such it
is a completely open problem.  This is the subject of this paper.

In the following we shall discuss what we believe to be the first attempt
to include higher nearest neighbour (displacements) shells in numerical
simulations of fractal growth in isotropic defect-free media of arbitrary
elastic constants.  We achieve this by (analytically and numerically) studying
Eq.(\ref{eq:w1}) under the condition that the Laplacian of $u$ is not
constant ({\em i.e.}, Laplace's case) or proportional to $u$ ({\em i.e.},
Poisson's case) along a growing surface.  Since this biharmonic operator
can also be generated in the theory of elasticity
\cite{Land59,Ball90,Meak89}, then ${\bf u}\equiv {\bf r}-{\bf r}'$
may become the displacement vector and the deformation of a body may be caused
by an {\em applied} force which appears in the solution of Eq.(\ref{eq:w1})
through some boundary conditions (Eq.(\ref{eq:w6}) below).

\section{Numerical simulation}

In this work, the numerical simulation of Eq.(\ref{eq:w1}) is carried out
on a lattice with either planar ({\em i.e.}, growth in a channel) or
circular ({\em i.e.}, radial growth) geometries.  Within the planar geometry
we use a $100\times 200$ lattice, having periodic boundary conditions along
the $x$-direction, and set the values $u^{o}=1$ and $u^{i}=0$ for the
upper $y$-boundary and the surface aggregate ($y=\ell$), respectively.
Initially, seed particles are placed on a line at displacement $u(i,1)$.
For circular geometry, we use lattice sites enclosed within a
circunference of radius $r=100$ and locate only one seed
at the center such that $u^{o}$ and $u^{i}$ are also unity and zero at
the outer (circular) boundary and at the inner (growing) aggregate,
respectively.

The inclusion of (second and higher-) nearest neighbour bond shells in
our numerical simulations follows the discrete form of Eq.(\ref{eq:w1})
on the ($i,j$) lattice site, which yields an expression involving
values of $u$ at {\em thirteen} mesh points written as \cite{Gde86}
\begin{eqnarray}\label{eq:w2}
& & u_{i-2,j}+2u_{i-1,j-1}-8u_{i-1,j}+2u_{i-1,j+1}+
  u_{i,j-2}-8u_{i,j-1}+20u_{i,j}  \\ \nonumber
  & & -8u_{i,j+1}+u_{i,j+2}+
2u_{i+1,j-1}-8u_{i+1,j}+2u_{i+1,j+1}+u_{i+2,j}=0   \;\;\; .
\end{eqnarray}
It is important to mention that numerical simulations using this equation
are somehow more involved than for Laplace \cite{Nie84} or Poisson
\cite{Lou92} growth because of long-range coupling.
However, the accuracy of the solution for Eq.(\ref{eq:w2}) can similarly
be improved by looking at the convergence of the
iterative solution using the Gauss-Seidel method \cite{Gde86}.

Equation (\ref{eq:w1}) requires modificiation
when applied at mesh points that are adjacent to a boundary, since one
(at the edge) or two (near the corners) of the values needed are at sites
outside the lattice.  This modification is made by introducing a fictitious
mesh point at ($i,L$) outside the planar lattice in the $y$-direction, where
the value of $u$ is there given by the derivative boundary condition
along one edge boundary:
\begin{equation}\label{eq:w6}
u_{i,L+1}=u_{i,L-1}+2h\frac{\partial u}{\partial y}\mid_{i,L}
    \;\;\; .
\end{equation}
Here $h\; (=h_{x}=h_{y})$ is the mesh size which we set equal to unity for
simplicity.  For planar geometry, we evaluate Eq.(\ref{eq:w6}) approximating
\begin{equation}\label{eq:madd1}
\frac{\partial u}{\partial y}\mid_{i,L}  \approx
   \frac{3(u^{o}-u^{i})}{L}\;\;\; ,
\end{equation}
for all $i$-columns.  This expression is obtained after taking the limit
$\ell /L <<1$ in the solution of a one-dimensional biharmonic equation.
Also for simplicity, the values of $u(i,0)$ for the points below the seed
will be set to zero.  For circular geometry we treat the boundary conditions
similarly to Laplacian growth \cite{Nie84}, {\em i.e.}, all the lattice sites
outside a circle of radius $r$ are set equal to $u^{o}$, hence, we approximate
for the sake of simplicity $\partial u/\partial y \mid_{y=L}\approx 0$
in the absence of applied field.

The procedure for growing our biharmonic aggregates follows standard
techniques \cite{Nie84}.  First, Eq.(\ref{eq:w1}) is solved iteratively
till the solutions converge to a given accuracity (of the order
$3\times 10^{-3}$ or smaller).  Second, after adopting a
growth probability law, the aggregates stochastically grow at one (or more)
perimeter(s) under a given relation between $u$ and growth probability $P$.

In order to have biharmonic pattern selection we shall assume that
$P$ at the grid site $(i,j)$ depends on two different phenomenological
(normalized) forms, namely
\[ P_{ij}= \left \{
  \begin{array}{ll}
    \frac{\mid \nabla ^{2}u_{i,j}\mid }{\sum \mid \nabla^{2}
        u_{ij}\mid} & \hspace{1cm} ({\rm model} \; I) \\
                      &  \\
    \frac{\mid u_{i,j}\mid }{\sum \mid u_{ij}\mid }  &
                      \hspace{1cm} ({\rm model} \; II) \;\;\; ,
   \end{array}
\right.  \]
where the sum runs over all of the nearest neighbour sites to an aggregagate.
Model $I$ implies that $P_{ij}$ is proportional to the local potential,
similarly to DBM
\cite{Nie84,Lou92,Meak91}. Whereas in model $II$, $P_{ij}$ is proportional
to displacements around the $(i,j)$ site.

\section{Results and discussion}

As already mentioned we attempt to study effects of long-range coupling
of displacements in simulations of fractal growth.  To achieve this, we start
analysing the numerical solutions of Eq.(\ref{eq:w1}) in planar geometry.
In Figs.1(a), (b) and (c) we show three characteristic stages of growing
for a biharmonic pattern by attaching one particle at each step
and assuming model $I$ for the probability $P_{ij}$.
Figure 1(a) includes $2100$ particles, (b) $4164$ particles and
(c) $4355$ particles added to the biharmonic cluster.
In these numerical simulations we estimate the derivative boundary condition
({\em c.f.}, Eq.(\ref{eq:w6})) from the analytical solution
given by Eq.(\ref{eq:madd1}) and fix $u^{i}=0$ and
(rescaled) $\phi^{o}\equiv \nabla^{2}u\mid_{y=L}\approx 6u^{o}/L^{2}$ such
that $u^{o}=1$.  Clearly, at a distance separation of about $\ell \sim 160$,
Fig.1(c) displays features of a transition from
dense to multibranched growth similarly to what occurs within Poisson growth
in planar geometry \cite{Lou92}.  This structure demostrates that long-range
coupling of displacements drives such phenomena.

Surprisely, such a transition also appears when we consider model $II$
for $P_{ij}$ as can be seen in Fig.2 for a biharmonic cluster containing
$3428$ particles.  This means that the influence of
ramified biharmonic patterns on the growth probability for each
lattice site is on the type of the (far from equilibrium) pattern obtained
and not on locating the transition.  We add that
several branches may also develop within our biharmonic growth
model by attaching -simultaneously, and stochastically- more than one
particle at each time step similarly to Louis {\em et al.}'s work
\cite{Lou92}.
We believe that for Poisson growth this transition is the result of
including many-body contributions via screening in a sort of mean field
approach. In our biharmonic model, long range coupling appears naturally as
a consequence of dealing with Eq.(\ref{eq:w1}).  As we shall show next
for the case of planar growth, the transition point within both Poisson and
biharmonic models appears when the growth velocities exhibit a minimum.

In Fig.3 we plot results for the grow velocity $v$ along the $y$-direction
for biharmonic growth under model $I$ (curve A), {\em i.e.}, proportional to
the averaged value of $P$ equal to $\nabla^{2}u$ to consider explicitly
nearest neighbour sites, and under model $II$ (curve B), where it has been
set proportional to the local displacement $u$.
For comparison, we also include in this figure results for
Poisson growth: {\em i.e.}, $\nabla^{2}\phi =\mu \phi$ \cite{Lou92}, such that
$v$ is set proportional to the field $\mid \phi_{i,j}-\phi^{i}\mid$
following Ref.\cite{Lou92}.  The Poisson growth model
incorporates screening (curve C): {\em i.e.}, $\mu =\lambda^{2}$ and
antiscreening (curve D): {\em i.e.}, $\mu =-\lambda^{2}$.

In this plot it can be seen that the transition from dense to
multibranched growth coincides with the fact that $v$ on the growing
surface presents a minimum.  This was first pointed out
by Louis {\em et al.} \cite{Lou92} for the case of screening (curve C)
(which also applies for antiscreening as seen in curve D).
However, from the present findings, we can argue that this is also true for
biharmonic growth (curves A and B) independently of how we have related
the probability $P_{ij}$ to $u(i,j)$.  For Laplacian growth
this phenomenon does not apper since the trend is to generate a single
tip at faster velocity than in the cases of Poisson or biharmonic growth.
It is worthwhile to point out that biharmonic patterns below the transition
point, {\em i.e.}, within the dense region (see Figs. 1 and 2), become
not that denser as for screening \cite{Lou92} but are rather denser
than for antiscreening (not shown).  Indeed, this can easily be understood
from Fig.3 since for Poisson growth (screening), such that $\mu >0$, the
transition occurs at smaller velocities than for biharmonic growth.
Henceforth, a Eden-like pattern can be generated due to screening
\cite{Lou92}.  Above the transition multibranched fractals appear in all
models considered, but within our biharmonic model the transition point
depends on the system size only (as it will become clear below).  We remark
that both Poisson and biharmonic models lead to obtain
growth velocities along the $y$-direction with parallel slopes
above and below their respective transition points.  We feel this to be a
particular feature of fractal growth due to long-range coupling of
displacements.

Let us focus next on effects due to long-range coupling on the
fractal growth in circular geometry.  This will allows us to make
calculations of the fractal dimension for the biharmonic patterns
by the standard box counting method \cite{Can91}.
To generate and select biharmonic fractal structures we set, for simplicity,
the derivative boundary condition required along the $r$-direction equal to
zero.  Figures 4(a) and (b) show the results for circular biharmonic growth
using again models $I$ (where $P$ is set proportional to
$\nabla^{2} u$, which corresponds to the potential in \cite{Lou92})
and $II$, respectively.  Figure 4(a) includes $2326$ particles
whereas Fig.4(b) $2172$ particles.
It can immediately be seen that above a certain transition point, say
$r_{\ell}$, these new plots give another
demostration that long-range coupling is the most relevant aspect for the
transition indepedently of the geometry adopted or of the given relation
between $P_{ij}$ and $u(i,j)$.

The transition point $r_{\ell}$ can, to a good approximation, be estimated
from the continuous limit of Eq.(\ref{eq:w1}) in cylindrical coordinates,
such that $z=0$ for all polar angles $\theta$, namely
\begin{equation}
\frac{1}{r^{2}}(\frac{\partial u}{\partial r}) - \;
  \frac{1}{r}(\frac{\partial^{2} u}{\partial r^{2}})+
      2(\frac{\partial^{3} u}{\partial r^{3}})+
           r(\frac{\partial^{4} u}{\partial r^{4}}) = 0  \;\;\; ,
\end{equation}
whose solution (far from the origin) is
\begin{equation}
u=A+Br^{2}+C\log r + Dr^{2}\log r  \;\;\; .
\end{equation}
Clearly, two of the four coefficients $A, B, C$ and $D$ may be determined
by the boundary conditions of the problem, {\em i.e.}, by assuming that at
$r=r_{\ell}$: $u(r_{\ell})=u^{i}\equiv 0$, whereas at $r=L$,
$u(L)=u^{o}\equiv 1$.  However, as discussed above, in addition to this
we also have the condition required by simulations
({\em c.f.}, Eq.(\ref{eq:w6})):
$\partial u/\partial r \mid_{r=L} \equiv 0$, which leads to
the following relation between $B, C$ and $D$:
\begin{equation}\label{eq:Madd4}
\frac{B}{D}=-\;\frac{C/D}{2L^{2}}-(\log L + \frac{1}{2})  \;\;\; .
\end{equation}
This in conjuction with one possible condition of stability for the
trajectories at the growing surfaces,
that is obtained from $\partial u/\partial r \mid_{r=r_{\ell}}\sim 0$:
{\em i.e.}, $B/D=-(\log r_{\ell} +1)$ and $C=0$, gives
\begin{equation}\label{eq:Madd2}
r_{\ell} \approx \frac{L}{e^{1/2}} = 0.607 \times L  \;\;\; .
\end{equation}
Hence, we have that the transition point only depends on the system size $L$
and occurs approximately at a distance (in lattice units) about $60 \%$ far
from the seed particle as indicated by arrows in Figs.4(a) and (b).
Nicely, our simplest prediction is in accord with the numerical simulations.
But if the derivative boundary condition
$\partial u/\partial r \mid_{r=L}$
is a constant different from zero, then Eq.(\ref{eq:Madd4}) changes by a
factor which, together with the above stability conditions, implies a
shift of the transition length and a more complicated relation between
$r_{\ell}$ and $L$.

To end this section, we examine the fractal properties of the simulated
biharmonic
patterns displayed in Figs.4(a) and (b), by simply counting the number of
particles, $N(r)$, inside a circle of increasing radius $r$ (in lattice units)
around a seed particle at the origin until we reach a distance $r_{\ell}$
at which the transition appears ({\em c.f.}, Eq.(\ref{eq:Madd2})).  We then
plot it as a function of $r$ in a $log$-$log$ plot, as despicted in Fig.5, for
models $I$ (triangles) and $II$ (squares) which we compare to Laplacian
growth (circles).  Over a decade, we obtain lines with slopes
larger than unity and smaller than the space dimension.  Thus, to a first
approximation, the fractal dimension of our biharmonic patterns in the denser
region approaches the value due to DBM (and DLA) within error bars
\cite{Nie84}.  This illustrates the fractal nature of the biharmonic patters.
Above $r_{\ell}$, there is a change of slope due to the transition
from dense to multibranched growth on contrary to the curve from
the Laplacian model (circles) which continues to be linear
indicating thus that the cluster grows dense.

\section{Concluding remarks}

In this work, we have shown that the discretization of the
biharmonic Eq.(\ref{eq:w1}) allows to reproduce a transition from dense
to multibranched growth at a point in which the growth velocity $v$
along the $y$-direction exhibits a minimum similarly to what occurs within
Poisson growth in planar geometry.  The physical basis for $v$,
as plotted in Fig.3, follows similarly to the dielectric breakdown model
(model $I$: curve A), namely assuming $v$ to be proportional to
the averaged value of $P$ equal to the potential $\nabla^{2}u$ -to
thus consider explicitly nearest neighbour sites.  Whereas, model $II$:
curve B, where $v$ has been set proportional to the local displacement
$u$, is simply a mathematical model.  On the other hand, we also discussed
that in circular geometry the transition point can be estimated
from the relation $r_{\ell}\approx L/e^{1/2}$ such that the trajectories
become stable at the growing surfaces in a continuous limit.  This
is in reasonable agreement with present numerical simulations.  Hence, we
conclude that the transition from dense to multibranched
growth within a biharmonic approach depends on the system size $L$ only.

The transition obtained from numerically solving the biharmonic equation
might not be necessarily similar to the novel and intriguing Hecker transition
(see {\em e.g.} \cite{Fleu91,Fleu92}).  But, to this end, we notice that when
assuming in the Kuramoto-Sivashinsky relation (\ref{eq:1q}), $\nu <0$ and
$\lambda >0$, then we obtain (see also \cite{Kura78})
\begin{equation}
\frac{\partial {\bf v} }{\partial t}=
    - \nabla^{2} {\bf v} + \nabla^{2}\nabla^{2}
          {\bf v}+2 {\bf v} \cdot \nabla  {\bf v}  \;\;\; ,
\end{equation}
in which ${\bf v}= \nabla u$ and $Curl \;{\bf v}=0$.  This looks somewhat
like the Navier-Stokes equation (for a potential flow with negative viscosity)
which may be somehow related to the recent analysis in Ref.\cite{Fleu92}
of electrochemical deposition.

Finally, it is important to emphasize again that -as a crucial difference to
those models satisfying second order differential equations- to select fractal
structures via the biharmonic part of Eq.(\ref{eq:1q}) requires values for
the order parameters (at either their first or second normal derivatives
at each boundary point) and the (second and higher-) nearest neighbour
bond shells on the ($i,j$) lattice site which yields an expression involving
values of $u$ at {\em thirteen} mesh points.  Thus, the formation of connected
patterns within the biharmonic equation is not trivial at all.
Because of this, our numerical simulations are more involved than for
Laplace (or Poisson) fractal growth due to the {\em long-range coupling}
between diplacements.

\vspace{2cm}

\section*{Acknowledgments}

The authors would like to thank the Condensed Matter Group, the
Computer Section and Prof. Abdus Salam at ICTP, Trieste, for support.

\newpage

\section*{Figure captions}

\begin{itemize}

\item {\bf Figure 1}:  Three characteristic stages of growing
for a biharmonic pattern by attaching one particle at each step
and assuming model $I$ for the probability $P_{ij}$:
(a) $2100$ (b) $4164$ and (c) $4355$ added particles to
diplay the transition from dense to multibranched growth.

\item {\bf Figure 2}:  Final biharmonic pattern above the transition
when considering model $II$ for $P_{ij}$ and $3428$ particles.

\item {\bf Figure 3}:  Growth velocity $v$ along the $y$-direction proportional
to $\mid \nabla^{2}u \mid$ (curve A) and displacement $\mid u \mid$ (curve B).
Curves C and D are for Poisson growth including screening and antiscreening,
respectively.

\item {\bf Figure 4}:  Results for circular biharmonic growth using
(a) model $I$ with $2326$ particles and (b) model $II$ with $2172$ particles.
Arrows indicate predicted transition points from Eq.(\ref{eq:Madd2}).

\item {\bf Figure 5}:  Fractal nature of the simulated biharmonic
patterns that are displayed in Figs.4(a) and (b).  Triangles are
obtained using model $I$ and squares using model $II$.  Circles are for
Laplacian growth giving a fractal dimension of $\sim 1.7$.

\end{itemize}

\end{document}